\documentclass[twocolumn,showpacs,preprintnumbers,amsmath,amssymb]{revtex4}
\usepackage{graphicx}

\newcommand{\half}{\mbox{\small $1 \over 2$}}

\def\bea{\begin{eqnarray}}
\def\eea{\end{eqnarray}}
\def\om{\omega}

\def\la{\langle}
\def\ra{\rangle}
\def\nn{\nonumber}
\def\cG{\mathcal{G}}
\begin{document}
\title{Linear response formula for finite frequency thermal conductance of open systems}
\author{Abhishek Dhar$^1$, Onuttom Narayan$^2$, Anupam Kundu$^1$ and Keiji Saito$^3$.}
\affiliation{$^1$ Raman Research Institute, Bangalore 560080, India}
\affiliation{$^2$ Department of Physics, University of California, Santa Cruz, CA 95064}
\affiliation{$^3$ Department of Physics, Graduate
  School of Science, University of Tokyo, Tokyo 113-0033, Japan.} 
\date{\today}
\begin{abstract}
An exact  linear response expression is obtained
for the heat current in a classical Hamiltonian system coupled to
heat baths with time-dependent temperatures.  The expression is
equally valid at zero and finite frequencies. We present numerical
results on the frequency dependence of the response function for
three different one-dimensional models of coupled oscillators
connected to Langevin baths with oscillating temperatures. For
momentum conserving systems, a low frequency peak is seen that,  
is higher than the zero frequency response for large systems. For
momentum non-conserving systems, there is no low frequency peak.
The momentum non-conserving system is expected to satisfy Fourier's law,
however, at the single bond level, we do not see any clear agreement with the 
predictions of the diffusion equation even at low frequencies. 
We also derive an exact analytical expression for the response 
of a chain of harmonic oscillators to a (not necessarily small) temperature
difference; the agreement with the linear response simulation  results for the
same system is excellent.
\end{abstract}

\maketitle

\section{Introduction}
\label{sec:intro}
In many low dimensional systems, heat transport unexpectedly violates
Fourier's law of heat conduction~\cite{lebowitzreview,LLPreview,dhar08}. This
can 
be because of integrability or proximity to integrability, which
is more common in low dimensions, as recognized starting from the
Fermi-Pasta-Ulam (FPU) model~\cite{FPU}. Alternatively, even  ergodic low-dimensional systems
can show anomalous heat conduction, with the conductivity diverging with system
size, if they conserve momentum. Apart from
the theoretical interest, understanding heat transport in such
systems is of relevance to heat conduction in carbon
nanotubes~\cite{expts}.

Most of the recent activity~\cite{LLPreview,dhar08} in this
field has dealt with the zero-frequency conductivity. But time
dependent temperature sources have been discussed in experimental
situations in the context of measuring the frequency dependent
thermal conductivity~\cite{dixon88,olafsen99} and specific heat~\cite{birge86}
of glassy systems. Theoretically, there have been a few studies on
the frequency dependent thermal current response using a microscopic
approach based on Luttinger's derivation of the Green-Kubo formula
and a hypothesis about the equality of certain transport
coefficients~\cite{bss06}, and from a phenomenological approach
\cite{volz01}. A recent paper studied thermal ratchet effects in
an inhomogeneous anharmonic chain coupled to baths with time-dependent
temperatures \cite{LHL08,ren10}.

In this paper, we adopt a different approach: we find the linear
heat conductance of a system placed in contact with two heat
reservoirs with time-dependent temperatures.  Physically the notion
of bath temperatures oscillating in time make sense if we assume
that the frequency of oscillation is much smaller compared to time
scales for local thermal equilibration in the reservoirs. An exact
expression (in the linear response regime) for the
heat current due to a small oscillating temperature difference
between the reservoirs is obtained.  

Our earlier result~\cite{kundu09} obtained the zero frequency
conductance of a finite system rather than the conductivity in the
infinite system limit.  Thus the thermodynamic limit was not taken
first (in fact, not at all), in contrast to the standard Green-Kubo
formula~\cite{greenkubo},  which cannot be applied when the infinite
system conductivity diverges. Our expression for the zero frequency
conductance involved the heat current auto-correlation function for
an open system. 
The extension to finite frequencies in this paper follows
the same approach, with the response now depending on the position
inside the system where the current is measured. 

We also show results of numerical simulations for the frequency dependent
response function by measuring the appropriate correlation function.
For one-dimensional momentum conserving anharmonic crystals, we
find a resonant response at a frequency $\omega\sim 1/N$ for a chain
of $N$ particles due to sound waves propagating from one end of the
system to the other. As $N$ increases, the resonance gets broader
and its height decreases slightly. However, its height relative to
the zero-frequency response {\it increases\/}, and for large $N$
this resonance is stronger than the zero frequency response.

We find that the low frequency peak disappears for systems where
momentum is not conserved. Fourier's law is known to be valid for
such systems, so that the heat current should satisfy the diffusion
equation. If one compares the numerical results for the frequency-dependent
heat current with the prediction from the diffusion equation at 
the single bond level, there seem to be substantial discrepancies. 

Numerical simulations for the frequency dependent response function
of a one-dimensional harmonic crystal, and an exact analytical
expression for the full  response (for finite $\Delta T$) of the same, are
also presented.  As far as we are aware of this is the first example of a
case where an analytical expression for the response
function has 
been obtained. For a harmonic system the full
response is also linear and hence we expect the linear response result to
agree with the exact response function. Indeed we find excellent agreement
between the numerical simulations of the  expression of the linear response
and the numerically evaluated exact  response expression. 

All three systems mentioned above also show a high-frequency peak
in the response function, whose location is independent of $N.$ One
can loosely ascribe this to the fact that the dynamics in the
interior of the system are underdamped (actually, undamped), so
that particles approaching each other recoil, and the heat current
auto-correlation function shows rapid oscillations in the temporal
domain. Such high-frequency oscillations are not seen in hard
particle models, such as the Random Collision Model~\cite{deutsch}.
This is discussed further when we derive the analytical expression for
the harmonic oscillator.
However, a quantitative understanding of the high-frequency peak
is lacking.

\section{Oscillator chains with Langevin baths}
\label{sec:whitenoise}
We follow the derivation of Ref.~\cite{kundu09}  to obtain  the
finite frequency heat conductance of an oscillator chain with
Langevin baths at the ends; more detail is provided in Ref.~\cite{kundu09}.
Consider the motion of $N$ particles on a one dimensional lattice,
described by the following Hamiltonian:
\begin{equation}
H = {1\over 2}\sum_{l=1}^N m_l v_l^2 + \sum_{l=1}^N U(x_l - x_{l+1}) 
+ \sum_{l=1}^N V(x_l)
\end{equation}
where ${\bf x} = \{x_l\}$ and ${\bf v} = \{v_l\}$ with $l=1,2,\ldots N$
are the displacements of the particles about their equilibrium positions
and their velocities, and $\{m_l\}$ are their masses.  We assume fixed
boundary conditions, $x_0 = x_{N+1} = 0.$ The particles 1 and $N$
are connected to white noise Langevin heat baths at temperatures $T_L$
and $T_R.$ Thus the equations of motion are
\begin{eqnarray}
m_l \dot v_l &=& 
-\frac{\partial}{\partial x_l} [U(x_{l-1} - x_l) + U(x_l - x_{l+1}) + V(x_l)]\nonumber\\
&+& \delta_{l,1}[\eta_L(t) -\gamma_L v_1] + \delta_{l,N}[\eta_R(t) -\gamma_R v_N]
\label{dyneq}
\end{eqnarray}
for $l = 1, 2,\ldots N.$ Here $\eta_{L,R}(t)$ are uncorrelated zero mean
Gaussian noise terms satisfying the fluctuation dissipation relations
\begin{equation}
\langle \eta_{L,R}(t)\eta_{L,R}(t^\prime)\rangle_\eta = 
2 \gamma_{L,R} k_B T_{L,R}\delta(t-t^\prime)~,
\end{equation}
where $\la ...\ra_\eta$ denotes an average over the noise. 

The derivation of the linear response theory starts with the Fokker-Planck
equation for the full phase space distribution function $P({\bf x}; {\bf v};
t)$.  If $T_L = T_R = T,$ the steady state solution to the equation is the 
equilibrium Boltzmann distribution. We now assume that the temperatures at the
two ends are oscillating in time with $T_{L,R} = T \pm \Delta T(t)/2$.
We will obtain a perturbative solution about the equilibrium solution.
The steps are very similar to the standard derivation of the fluctuation
dissipation theorem.  The Fokker Planck equation corresponding to
Eq.~(\ref{dyneq}) is
\begin{equation}
\frac{\partial P}{\partial t} = 
-\sum_l \frac{\partial}{\partial x_l} (v_l P) 
- \sum_l\frac{\partial}{\partial v_l} (f_l P/m_l)
+ O_1 P + O_N P
\label{fp}
\end{equation}
where $f_l=-\partial H/\partial x_l$ is the force acting on the $l$'th particle.  The operators
$O_{1,N}$ come from the Langevin damping and noise on the terminal
particles:
\begin{eqnarray}
O_1 P &=& \frac{\gamma_L}{m_1} \frac{\partial}{\partial v_1} (v_1 P)
+ \frac{\gamma_L k_B T_L}{m_1^2} \frac{\partial^2}{\partial v_1^2} P
\nonumber\\
O_N P &=& \frac{\gamma_R}{m_N} \frac{\partial}{\partial v_N} (v_N P)
+ \frac{\gamma_R k_B T_R}{m_N^2} \frac{\partial^2}{\partial v_N^2} P.
\label{oj}
\end{eqnarray}
With $T_{L,R} = T \pm \Delta T(t)/2,$ we can group terms according to their
power of $\Delta T$ to obtain
\begin{equation}
\frac{\partial P}{\partial t} = 
= \hat L P + \hat L^{\Delta T} P
\end{equation}
where 
\begin{equation}
\hat L^{\Delta T} = 
\frac{k_B\Delta T}{2} \bigg[\frac{\gamma_L}{m_1^2}\frac{\partial^2}{\partial v_1^2} 
- \frac{\gamma_R}{m_N^2}\frac{\partial^2}{\partial v_ N^2}\bigg].
\label{LDeltaT}
\end{equation}
For $\Delta T = 0,$ the steady state solution of the Fokker Planck
equation is the equilibrium Boltzmann distribution $P_0 = \exp[-\beta
H]/Z,$ where $Z$ is the canonical partition function and $\beta = 1/(k_B
T).$ For $\Delta T\neq 0,$ we start with the equilibrium distribution at
time $t=t_0$ and then let the system evolve under the full Fokker Planck
operator. Writing $P({\bf x}, {\bf v}, t) = P_0 + p({\bf x}, {\bf v},
t)$ and retaining terms to $O(\Delta T),$
\begin{equation}
\frac{\partial p}{\partial t} = \hat L p + \hat L^{\Delta T} P_0.
\end{equation}
Setting $t_0\rightarrow -\infty$ we get the formal solution to this
equation
\begin{equation}
p({\bf x};{\bf v}; t) = 
\int_{-\infty}^t e^{(t - t^\prime)\hat L} ~\Delta\beta(t')~J_{fp}({\bf v}) P_0({\bf x}, {\bf v}) dt^\prime
\end{equation}
where $J_{fp}({\bf v})$ is defined by 
\begin{equation}
\frac{\partial P}{\partial t}\bigg\vert_{P=P_0} = \hat L^{\Delta T} P_0 =
(\Delta\beta) J_{fp} P_0 
\label{jfpdef}
\end{equation}
from which
\begin{equation}
J_{fp}  = {\gamma_R\over {2 m_N}} [m_N v_N^2 - k_B T]
- \frac{\gamma_L}{2 m_1}[m_1 v_1^2 - k_B T].
\label{jfpdef1}
\end{equation}

The expectation value of any function $\langle \Delta A\rangle = \langle
A\rangle -\langle A\rangle_0$ of any observable $A({\bf x}; {\bf v})$
then takes the form:
\begin{equation}
\langle \Delta A(t)\rangle_{\Delta T} = -\frac{1}{k_B T^2}
\int_0^\infty \langle A(\tau) J_{fp}(0)\rangle \Delta T(t-\tau) d\tau
\label{expvalff}
\end{equation}
where we have defined the equilibrium average $\la A(t) J_{fp}(0)\ra = \int
d {\bf x} \int d {\bf v} A e^{\hat{L} t} J_{fp} P_0$ and we have used the time
translational invariance of the equilibrium 
correlation function. In particular, we are interested in the
energy current between two adjacent particles~.  The instantaneous
current from the $l$'th to the $l+1$'th site is given by: $ j_{l+1,l}
= \half (v_l + v_{l+1}) f_{l+1,l}$, where $f_{l+1,l}=-\partial
U(x_l-x_{l+1})/\partial x_{l+1}$ is the force on the
$l+1$'th particle due to the $l$'th particle.  We  get for the
average heat current flowing between any bond on the chain  by:
\begin{equation}
\langle j_{l+1,l}(t)\rangle_{\Delta T} = -\frac{1}{k_B T^2}
\int_0^\infty \langle j_{l+1,l}(\tau) J_{fp}(0)\rangle \Delta T(t-\tau) d\tau.
\label{j12ff}
\end{equation}
For a oscillating temperature given by $\Delta T(t)=\Delta T(\omega)
e^{i\omega t}$ this gives:
\begin{eqnarray}
\frac{\langle j_{l+1,l}(\omega)\rangle}{\Delta T(\omega) e^{i \omega
    t}} = G_l(\omega) e^{-i \phi_l(\omega)}~~~~~~~~~~~~~~\nn\\
~~~~~= -\frac{1}{k_B T^2}
\int_0^\infty \langle j_{l+1,l}(\tau) J_{fp}(0)\rangle e^{-i \omega
    \tau} d\tau~,~~  
\label{result}
\end{eqnarray}
where $G_l(\omega)$ is the magnitude of the response --- to be
computed numerically in Section~\ref{sec:numerical} --- and $\phi_l$
is the phase. The correlation function $\langle j_{l+1,l}(\tau)
J_{fp}(0)\rangle $ on the right hand side of this equation is for
a system in equilibrium at temperature $T.$

A few comments are appropriate here. First, as shown in Ref.~\cite{kundu09},
for $\omega\rightarrow 0$ it is possible to manipulate the integrand
on the right and make it proportional to the aut-correlation function
of the heat current integrated over the entire chain, $\sum_l
j_{l+1,l}(\tau),$ yielding a result resembling the standard Green-Kubo
formula (but without the thermodynamic limit). This manipulation
is not possible for $\omega \neq 0.$ Thus the current response depends on $l,$
the position inside 
the chain where the response is measured, as one would
expect. Moreover, the correlation function involves $J_{fp},$ which
is different from the heat current.

Second, although we have assumed that $\Delta T_L = -\Delta T_R$
to resemble the zero-frequency calculations of Ref.~\cite{kundu09}
where such an assumption is appropriate, at $\omega\neq 0$ there
is no reason why one cannot treat $\Delta T_L$ and $\Delta T_R$ as
independent variables.  It is straightforward to extend the derivation
above and obtain the response to $\Delta T_R$ and $\Delta T_L,$
with $J_{fp}$ in Eq.~(\ref{result}) replaced by the first and second
part of Eq.~(\ref{jfpdef1}) respectively. For large $N,$ one expects
that the response to a oscillatory temperature perturbation at one
end of the chain should only depend on the distance from that end
and be the same as for a semi-infinite chain.

Finally, expressions similar to Eq.~(\ref{j12ff}) can be obtained
for any quantity that depends on the phase space variables of the
system, not just $j_{l+1,l}(\tau).$ It does {\it not\/} apply to
the heat current flowing into the system from the reservoirs, since
they involve the Langevin noise $\eta_{L,R},$ and these have to be
obtained indirectly.  Thus Eq.~(\ref{j12ff}) is valid for $l=1,$ and
one also has
\begin{equation}
\langle d\epsilon_1(t)/dt\rangle_{\Delta T} = -\frac{1}{k_B T^2}\frac{d}{dt}
\int_0^\infty \langle \epsilon_1(\tau) J_{fp}(0)\rangle \Delta T(t-\tau) d\tau.
\label{e1ff}
\end{equation}
Replacing the $d/dt$ with a $-d/d\tau$ acting on $\Delta T$ and
integrating by parts, adding this to Eq.~(\ref{j12ff}), and using
the fact that $j_{21}(t) + d\epsilon_1(t)/dt = j_{1,L}(t)$ (where
$j_{1,L}$ is the heat current flowing in from the left reservoir),
we have
\begin{eqnarray}
\langle j_{1,L}(t)\rangle_{\Delta T} &=& -\frac{1}{k_B T^2}
\int_0^\infty \langle j_{1,L}(\tau) J_{fp}(0)\rangle
\Delta T(t-\tau) d\tau \nonumber\\
&-& \frac{1}{k_B T^2}\Delta T(t)\langle \epsilon_1(0) J_{fp}(0)\rangle.
\label{j1Lff}
\end{eqnarray}
Fourier transforming, for $\Delta T(t) = \Delta T(\omega) e^{i\omega t},$
the heat current flowing from the left reservoir is
\begin{equation}
\Bigg\langle \frac{j_{1,L}(\omega)}{\Delta T(\omega)}\Bigg\rangle = 
-\frac{1}{k_B T^2}\int_0^\infty \langle j_{1,L}(\tau) J_{fp}(0)\rangle 
e^{-i\omega \tau} d\tau + \frac{\gamma_L}{m_1} k_B.
\label{j1LffFT}
\end{equation}
This response function has a non-zero $\omega\rightarrow\infty$
limit from the second term on the right hand side. This is reasonable:
if $\Delta T$ oscillates at a very high frequency, the effect on
$({\bf x}, {\bf v})$ should be negligible, but the current flowing
from the left reservoir should oscillate because $\langle\eta_L(t)
v_1(t)\rangle_\eta = \gamma_L k_B T_L(t)/m_1$ is proportional to
the instantaneous temperature of the reservoir. The instantaneous
response of Eq.~(\ref{j1LffFT}) is a peculiarity of white noise
stochastic baths, and is not seen for Nose-Hoover baths --- where
even the heat current at the boundary is in terms of the extended
phase space variables --- or a fluid system with Maxwell boundary
conditions where continuity requires that the heat current at the
boundary and just inside the system should be the same. Therefore,
hereafter we work with $j_{21}$ and $j_{N,N-1}$ when we want the 
current at the boundaries.

Although the derivation given above is for a one-dimensional chain,
it is straightforward to see that it is valid for any system that
is connected to only two reservoirs, regardless of its dimensionality.

\section{Numerical Results}
\label{sec:numerical}
\begin{figure}{!htbp}
\vspace{1.0cm}
\includegraphics[width=3.3in]{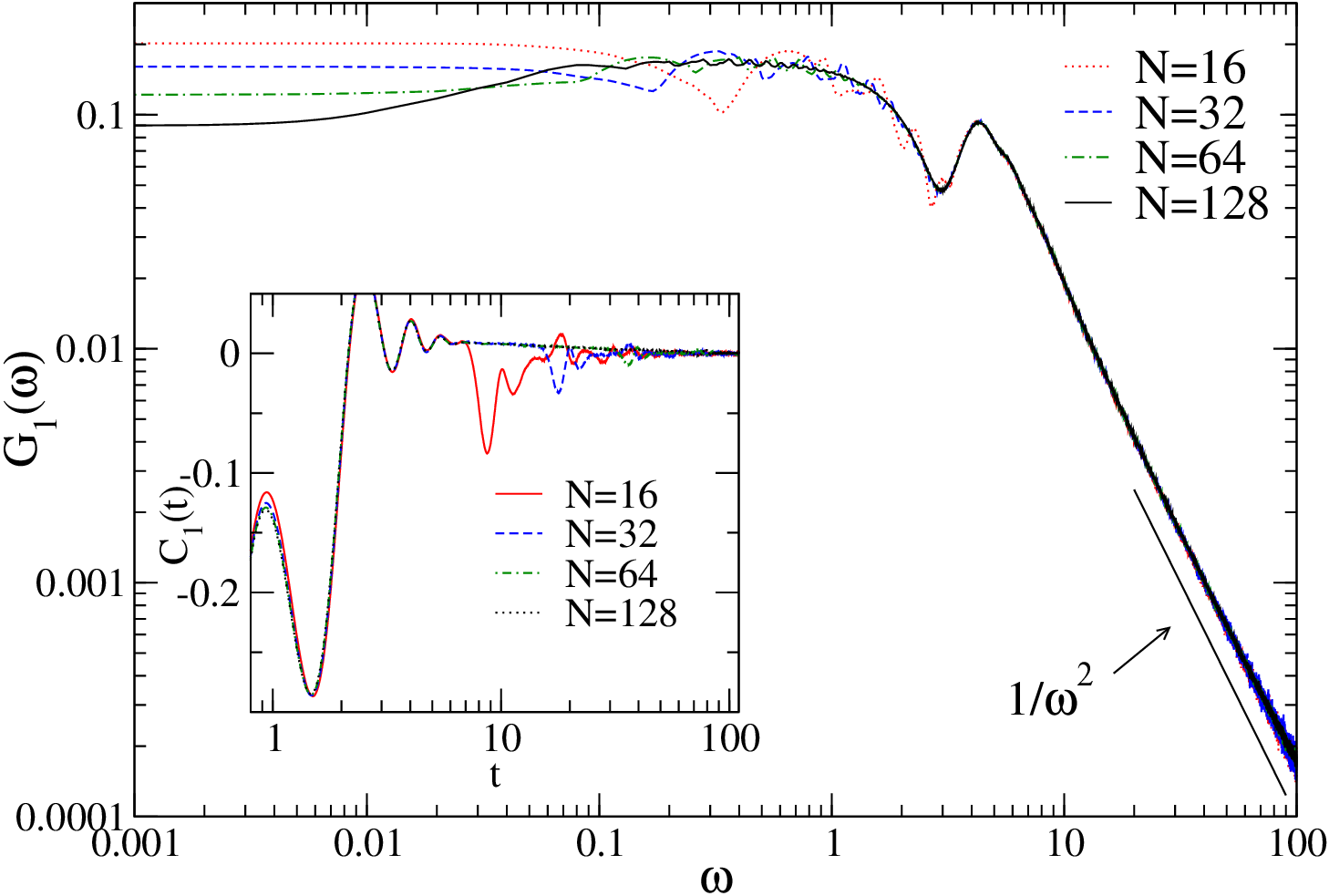}
\caption{(Color online) Plot of  magnitude of the response function, $G_1(\omega),$
for FPU chains of different lengths.
The inset shows the correlation
function $C_1(t),$ which has the same information in the time
domain.}
\label{fpucw_1}
\end{figure}

Numerical simulations to obtain the correlation function on the
right hand side of Eq.~(\ref{result}) were performed on three different
systems, which differ in the potential of each particle. From these correlation
functions we obtained $G_l(\omega)$ using Eq.~(\ref{result}). The
velocity-Verlet algorithm was used, with a time step $\delta t =
0.005.$ We verified that doubling $\delta t$ does not change our
results.
For the largest systems, the initial equilibration time was $t_{eq}
= 64\times 10^6,$ after which the dynamical equations were evolved
for a time $t = 5\times 10^8.$ All the particle masses were set to $1$,
$\gamma_L = \gamma_R = 1,$ and the reservoirs were at temperature
$T=2.0$. 
Figure~\ref{fpucw_1} shows $G_1(\omega)$ as a function of $\omega,$
as defined by Eq.~(\ref{result}), for FPU chains of different lengths.
The potential used was $U(x) = x^2/2 + x^4/4$ with $V(x) = 0.$ An
$N$-independent high frequency peak and a low frequency peak at
$\omega\sim 1/N$ are seen. Higher harmonics of the low frequency
peak can be barely discerned. As the system size is increased, the
low frequency peak broadens and decreases slightly in height, but
the zero frequency response drops much faster. Thus by $N=128,$ the
$\omega\sim 1/N$ resonance is clearly {\it stronger\/} than the
zero frequency response. Note that Eq.~(\ref{result}) gives the
conductance, not the conductivity; the $\omega = 0$ conductance
decreases as $\sim 1/N^{1-\alpha}$. It is expected that $\alpha = 1/3$
\cite{dhar08} for large $N$ but this  would require much larger system sizes to
verify.  
The inset to Figure~\ref{fpucw_1} shows $C_1(t) = \langle
j_{21}(t) J_{fp}(0)\rangle,$ i.e. the same information in the time
domain. $N$-independent short time oscillations that decay to
(approximately) zero are seen.  An `echo' of the oscillation is
seen at a time $\tau_N$ that is approximately $N/v,$ where $v$ is
possibly related to the velocity of effective phonons \cite{li10}.
At high frequencies, $G_1(\omega)$ is approximately independent of
$N$ as one would expect, with a high frequency peak.  As
$\omega\rightarrow\infty,$ $G_1(\omega)\sim 1/\omega^2.$

\begin{figure}
\vspace{1.0cm}
\includegraphics[width=3.3in]{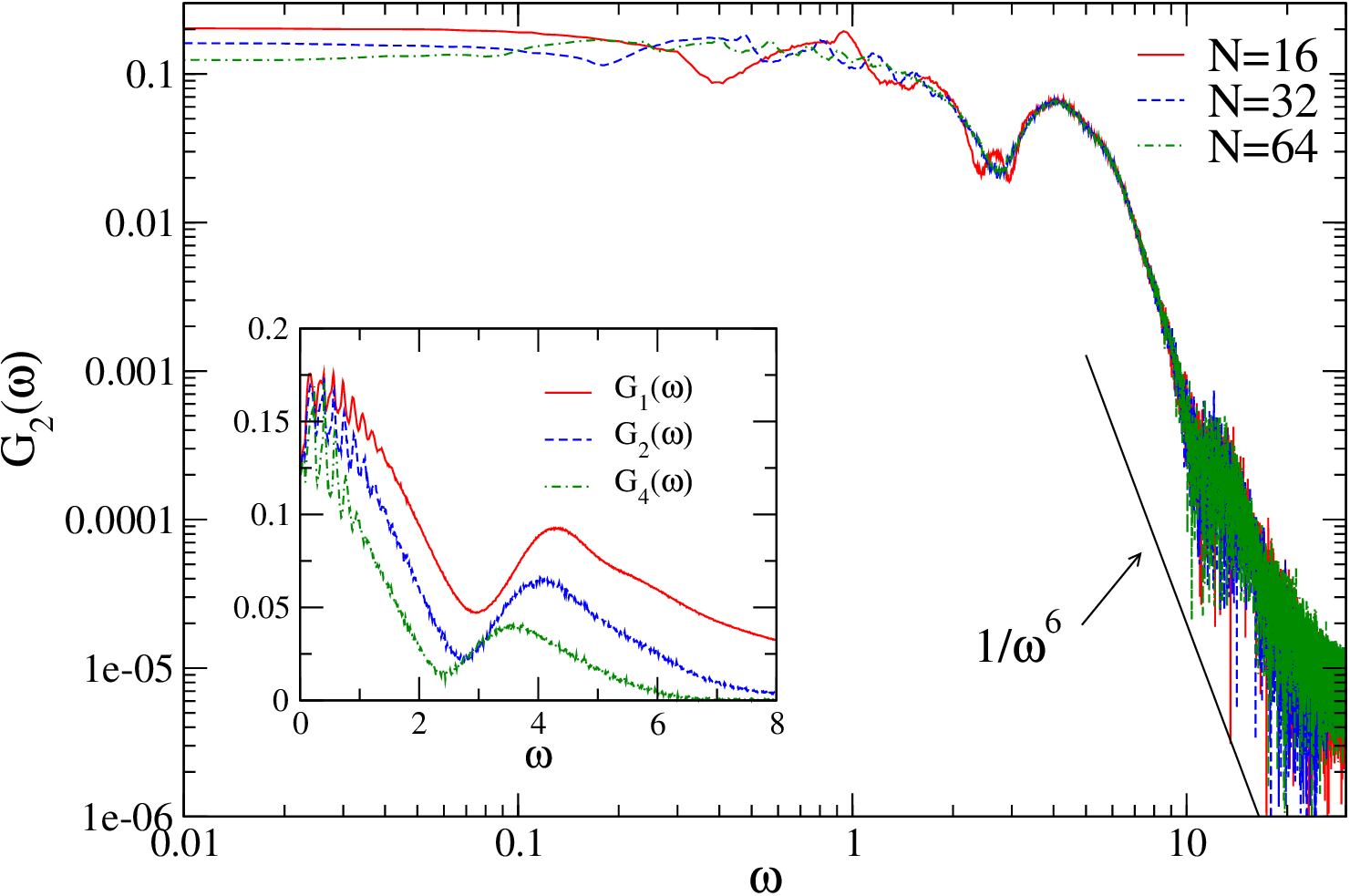}
\caption{(Color online) Plot of  the magnitude of the response function, $G_2(\omega),$
for FPU-chains of different lengths. The inset shows $G_l(\omega)$ for 
various $N=64$ and various $l.$}
\label{fpucw_x}
\end{figure}

Figure~\ref{fpucw_x} shows $G_2(\omega),$ the magnitude of the response function at
a distance $l=2$ from the left boundary. The low frequency peak
(and its harmonics) are still present, but much more irregular in
shape. However, from a device perspective,
it is the currents flowing into the boundaries that are
important.
The high frequency behavior is independent of $N,$ and as seen in
the inset, the peak in $G_l(\omega)$ shifts to smaller $\omega$ as
$l$ is increased. 
It is not clear if $G_2(\omega\sim \infty)\sim
1/\omega^6$ as is seen for the harmonic chain (discussed
later in this paper).

\begin{figure}
\vspace{1.0cm}
\includegraphics[width=3.3in]{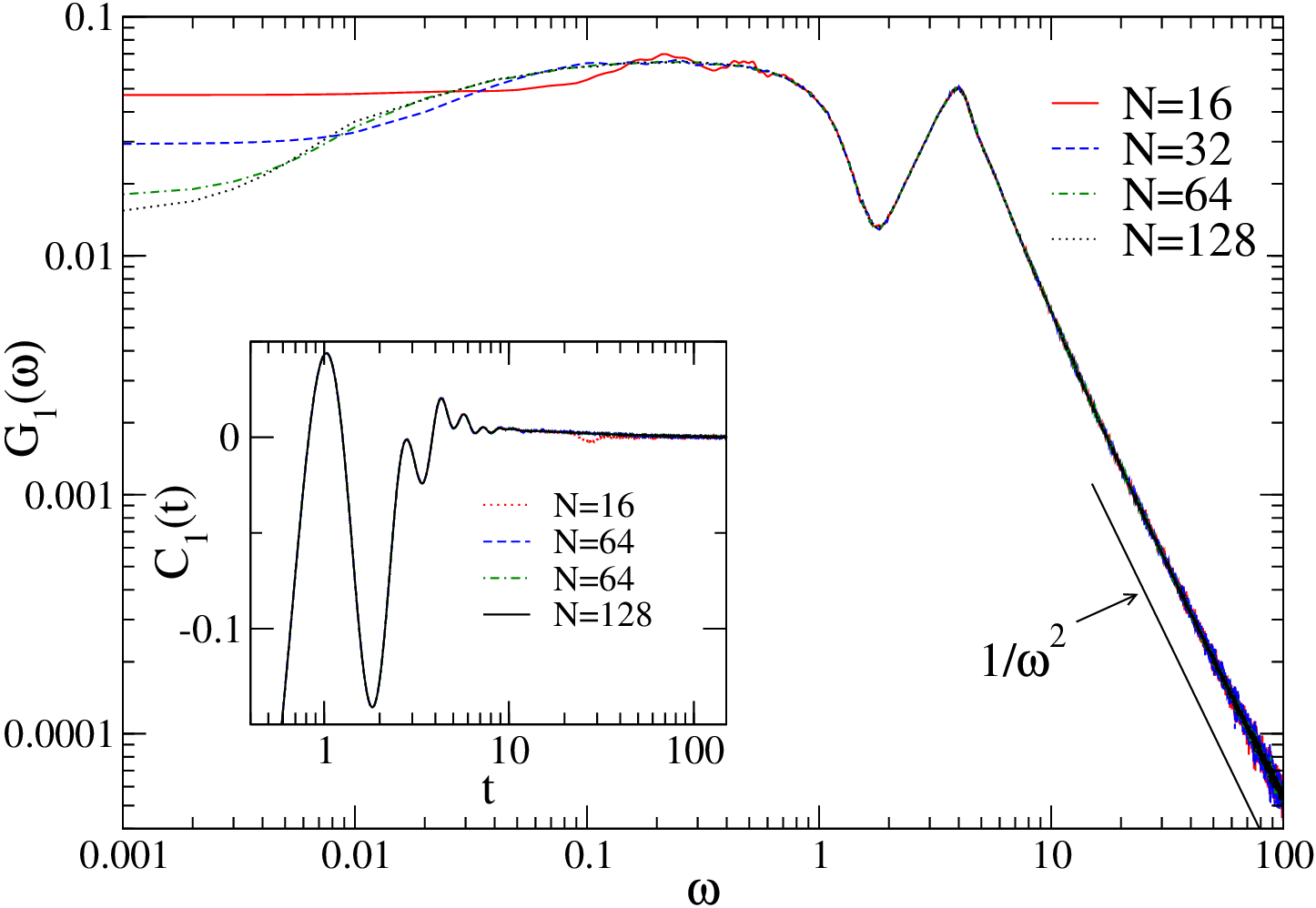}
\caption{(Color online) Plot of the magnitude of the response function, $G_1(\omega),$
for $\phi^4$-chains of different lengths.
The inset shows $C_1(t).$}
\label{phi4cw_1}
\end{figure}
\begin{figure}
\vspace{1.0cm}
\includegraphics[width=3.3in]{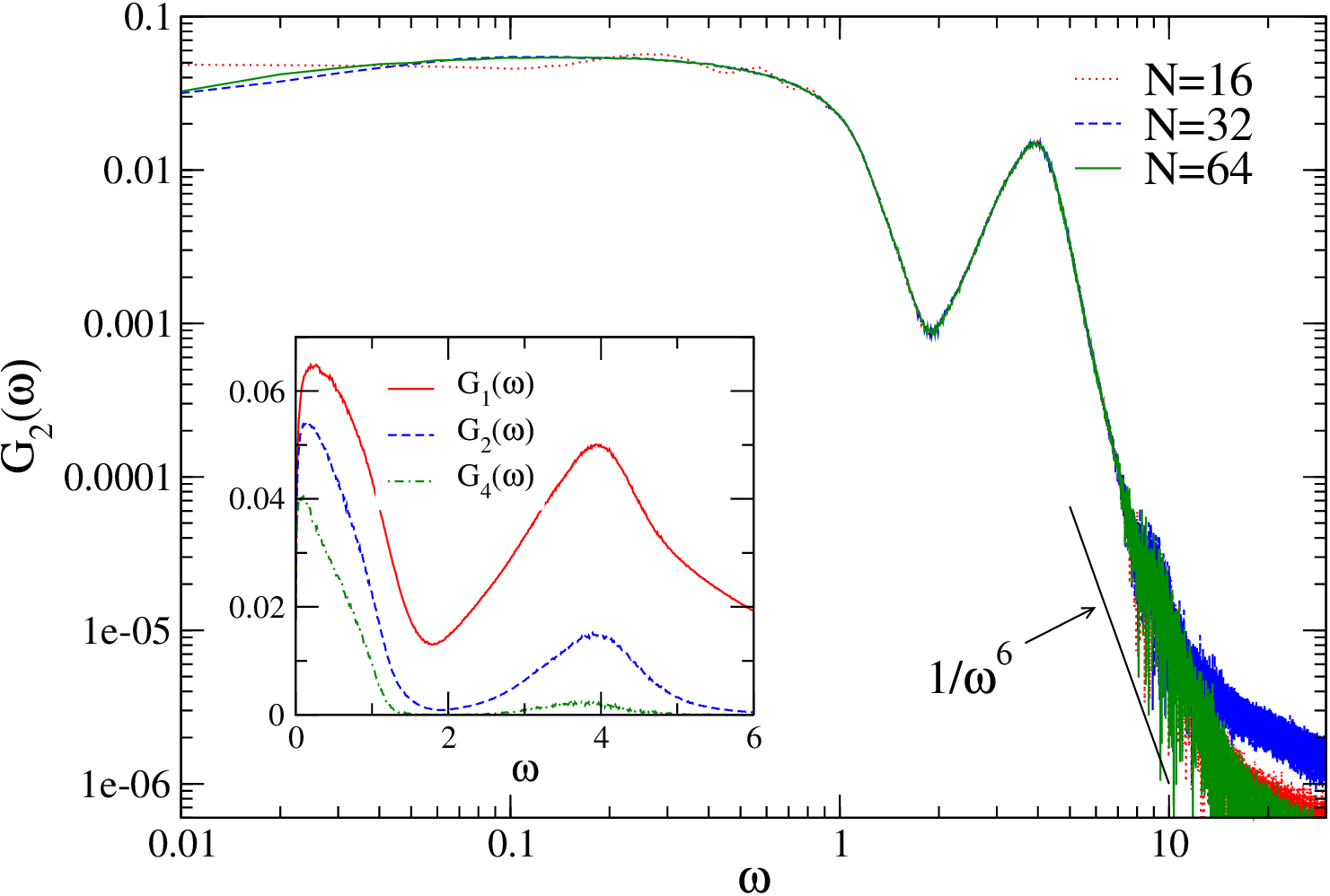}
\caption{(Color onine) Plot of  $G_2(\omega)$
for $\phi^4$-chains of different lengths. A fit to $\sim 1/\omega^6$
in the asymptotic high frequency regime is shown. The inset has 
$G_l(\omega)$ for various $l$ and $N=64.$}
\label{phi4cw_x}
\end{figure}

 
Figure~\ref{phi4cw_1} shows $G_1(\omega)$ for
chains of different lengths with an onsite potential $V(x) = x^4/4.$
The interparticle potential is harmonic, $U(x) = x^2/2.$ The dynamics
are not momentum conserving, and the zero frequency conductance
should be inversely proportional to $N.$ This is not seen in the
data for two reasons: direct measurement of the zero frequency conductance by
applying a small temperature difference between the reservoirs shows
that one needs $N\gtrsim 256$ to see the $\sim 1/N$ dependence, and the
curves for the two larger systems (more noticeably $N=128$) have not 
reached their $\omega\rightarrow 0$ limit in the figure.
The low frequency resonance is gone, replaced by a broad $N$-independent
plateau. This is presumably because at finite temperature, the
effective phonons are optical instead of acoustic.
The $N$-independent high frequency peak is also present. The response
in the interior of the chain, shown in Figure~\ref{phi4cw_x} is similar, 
except that the low frequency
plateau extends down to $\omega=0$ (or to very small $\omega$). As for the 
FPU chains, we fit $G_1(\omega\sim\infty$ to $\sim 1/\omega^2$ and 
--- less successfully ---
$G_2(\omega\rightarrow\infty)$ to $\sim 1/\omega^6.$  
From the inset to Figure~\ref{phi4cw_x}, there is no significant
$l$-dependence to the location of the high frequency peak in
$G_l(\omega)$, unlike what we saw for FPU chains.

\begin{figure}{!htbp}
\vspace{0.7cm}
\includegraphics[width=3.3in]{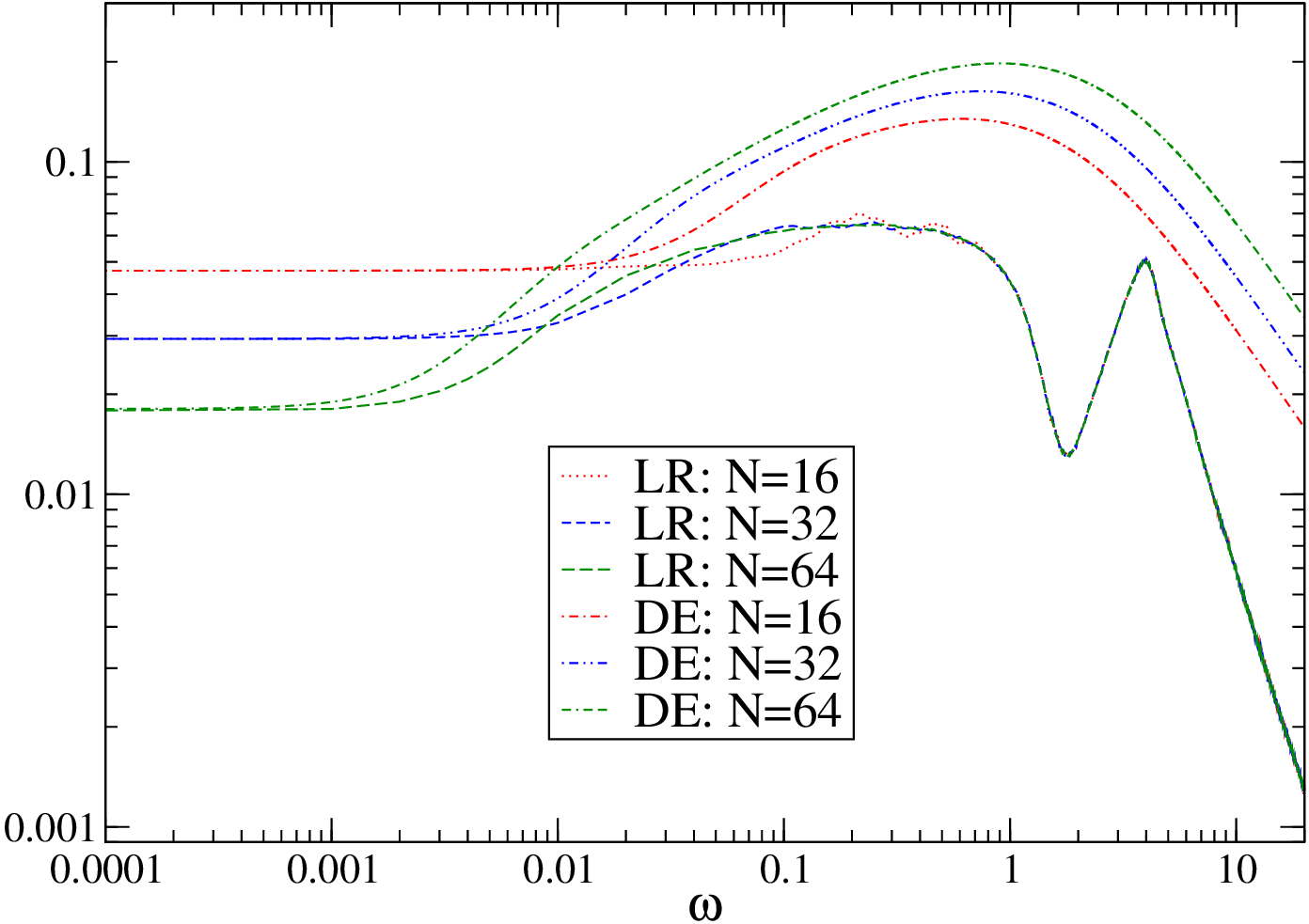}
\caption{(Color online) Plot of $G_1(\omega)$ for $\phi^4$ chains of different
lengths (LR) and $G^{\rm diff}_1(\om)$ from the diffusion equation (DE).}    
\label{compare}
\end{figure} 
Beyond the $\sim 1/N$ dependence of the zero frequency conductance,
one expects that heat transport in systems that are not momentum
conserving should be diffusive, and the temperature field will
satisfy $\partial T_l/\partial t = D (T_{l+1} - 2 T_l + T_{l-1})$
where $D = \kappa/C$ is the diffusion constant. With an $\sim
e^{i\omega t}$ time dependence, the resultant difference equation
can be solved with $T_L(\omega)$ and $T_R(\omega)$ specified, and
thence the heat current $j_{l+1,l} = \kappa (T_l - T_{l+1})$ can
be calculated. Some features of the solution are $G^{\rm diff}_l(\omega
= 0) \propto 1/N,$ $G^{\rm diff}_l(\omega)$ is independent of $N$ for 
$N\rightarrow\infty,$ $G^{\rm diff}_l(\omega\rightarrow 0) \sim \omega^{1/2}
\exp[-(\omega/2D)^{1/2} l]$ and $G^{\rm diff}_l(\omega\rightarrow\infty)\sim
1/\omega^l.$ In Figure~\ref{compare} we plot the responses $G^{\rm diff}_{1}$
together with the linear response results $G_{1}$ for the $\phi^4$ model.  
For each system size we fix the diffusion constant $D$ so that the $\om=0$
results for the two responses match.   One expects that the low-frequency
agreement between the two sets should become better with increasing system
size. However  this is not clear from our data.
At high frequencies, the expectation  $G^{diff}_1(\omega)\sim 1/\omega$ is 
definitely not borne out. Since the diffusion equation is not expected to be
valid at microscopic time or length scales, and the fact that $\sim 1/N$
scaling of the zero frequency heat conductance is only seen for $N\gtrsim
256$ suggests that `microscopic' length scales are quite large here, the
lack of agreement at the single bond level and high frequencies is perhaps not
surprising. A clear understanding of this requires further work.

\begin{figure}{!htbp}
\vspace{0.7cm}
\includegraphics[width=3.3in]{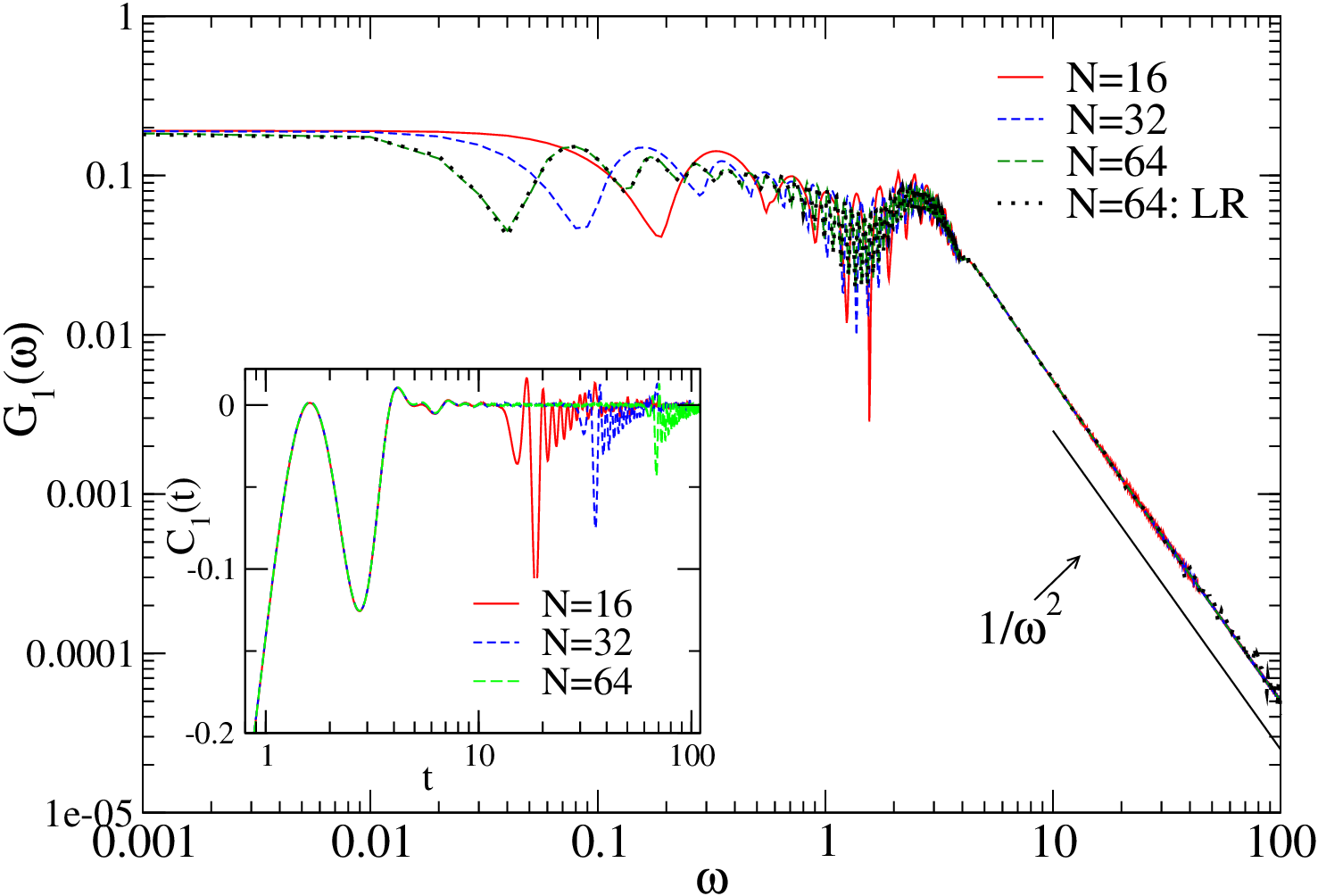}
\caption{(Color online) Plot of $G_1(\omega)$ for harmonic chains of different
lengths, from the analytical expression derived in Section~\ref{sec:anal}.
Because of the complicated structure in the figure, $N=128$ is not
included.  The linear response simulation results for $N=64$ are also shown
(LR). The inset shows $C_1(t)$.}
\label{harmcw_1}
\end{figure}
\begin{figure}
\vspace{1.0cm}
\includegraphics[width=3.3in]{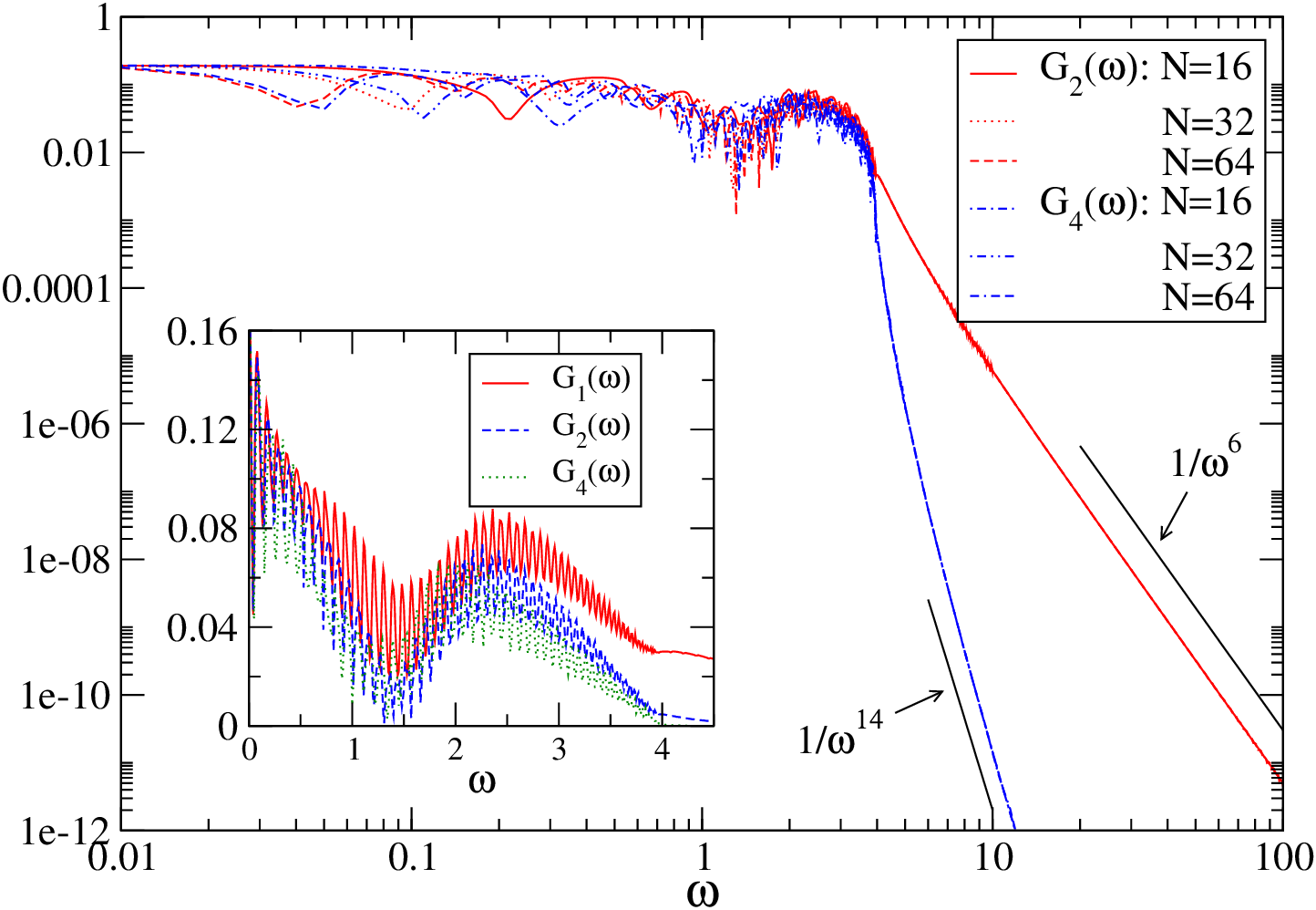}
\caption{(Color online) Plot of $G_2(\omega)$  and $G_4(\omega)$ for harmonic
chains of various lengths. The fit to the asymptotic form
$G_l(\omega\rightarrow\infty)\sim 1/\omega^{4 l - 2}$ is shown. The
inset shows $G_l(\omega)$ for various $l$ and $N=64.$}
\label{harmcw_x}
\end{figure}

 
Finally, we show the results for a harmonic chain, with $V(x) = 0$
and $U(x) = x^2/2.$  In this case we show in the next section
[sec.~(\ref{sec:anal})] that the response $G_l(\omega)$ can be obtained
exactly and expressed in terms of a  single integral over frequencies. 
Here we give numerical results for $G_l(\om)$ obtained using this exact
formula [Eq.~(\ref{exresp})] and also compare it with the linear response result  [Eq.~(\ref{result})].  
We show $G_1(\omega)$ in Figure~\ref{harmcw_1}, with 
results from numerical simulations of the linear response formula also shown
for $N=64$. We see excellent agreement between the analytical and linear
response result.
One can see that $G_1(\omega=0)$ is almost $N$-independent as expected,
and the low frequency resonance and its harmonics are more
pronounced than for the FPU chain, which is not surprising since
there is no dispersion or damping in the interior of the chain. The 
high frequency peak seems to be present but is difficult to cleanly
separate from the low frequency structure. As was the case for the FPU
and $\phi^4$ chains, 
$G_1(\omega\rightarrow\infty)$ is $N$-independent and $\sim 1/\omega^2.$ 
In Section~\ref{sec:anal}, the asymptotic form $G_l(\omega\rightarrow\infty)\sim 1/\omega^{4 l - 2}$ is obtained.
Figure~\ref{harmcw_x} shows $G_2(\omega)$ for various system sizes, 
with all features as expected. 

\section{Response of a harmonic chain}
\label{sec:anal}
Although the integrability of the harmonic oscillator chain makes
its behavior non-generic, and its applicability to physical systems
limited, the advantage of this model is that its response can be
completely obtained analytically (with some integrals evaluated
numerically) and compared to the simulation results.  We now proceed
with the analysis.

In this case both $V(x)$ and $U(x)$ are quadratic
and the Hamiltonian can be written in the form $H=\dot{X}M\dot{X}/2+X\Phi X/2$
where $M$ and $\Phi$ are respectively the mass matrix and the force-constant
matrix for the system.  
We will obtain the solution of the equations of motion in the 
time-dependent   steady state by using Fourier transforms in the time
domain. The approach is 
similar to that used in the derivation of the Landauer-type formula for steady
state heat current in harmonic systems, where the current is expressed in
terms of phonon Green's functions \cite{dharroy06}. 
Let us introduce
the transforms: $\tilde{x}_l(\Omega)=(1/2 \pi) \int_{-\infty}^{\infty} dt
x_l(t) e^{i \Omega t} $ and  $\tilde{\eta}_{L,R}(\Omega)=(1/2 \pi)
\int_{-\infty}^{\infty} dt \eta_{L,R}(t) e^{i \Omega t} $. Then the Fourier transform
solution of Eqns.~(\ref{dyneq})  gives:
\bea
\tilde{x}_l(\Omega) = \cG_{l1}^+(\Omega) \tilde{\eta}_L(\Omega)
+\cG_{lN}^+(\Omega) \tilde{\eta}_R(\Omega)~,   
\eea 
where $\cG^+(\Omega)=[-M \Omega^2 +\Phi -\Sigma^+(\Omega) ]^{-1}$ is the
phonon Green's function \cite{dharroy06} and $\Sigma^+$, the self-energy correction due to
baths,  is a $N
\times N$ matrix whose only non-zero elements are $\Sigma^+_{11}=i \Omega
\gamma_L, ~ \Sigma^+_{NN}=i \Omega \gamma_R$. The noise correlations
corresponding to the oscillating temperatures $T_L=T+\Delta T/2 \cos \om
t,~T_R=T-\Delta T/2 \cos \om t $ are given by: 
\bea
&&\la \tilde{\eta}_{L}(\Omega_1)   \tilde{\eta}_{L}(\Omega_2) \ra =
\frac{\gamma_{L} k_B }{\pi} \{~T~ \delta (\Omega_1+\Omega_2)  \nn \\ 
&+&(\Delta T/4) [~ \delta
  (\Omega_1+\Omega_2+ \om) +  \delta (\Omega_1+\Omega_2 -
  \om)~]~ \}~,    \nonumber \\
&&\la \tilde{\eta}_{R}(\Omega_1)   \tilde{\eta}_{R}(\Omega_2) \ra =
\frac{\gamma_{R} k_B }{\pi} \{~T \delta (\Omega_1+\Omega_2) \nn \\ &-& (\Delta T/4) [~\delta
  (\Omega_1+\Omega_2+ \om) + \delta (\Omega_1+\Omega_2 -
  \om)~] ~\}~,~~ \label{noisew}  
\eea
and $\tilde{\eta}_L, \tilde{\eta}_R$ are uncorrelated. 
The heat current on any bond is given by the noise average 
$\la j_{l+1,l}\ra= \la (1/2) \la k(x_l-x_{l+1}) (v_l+v_{l+1}) \ra,$ where $k$ is the 
force constant of the bonds, and thus
involves evaluating 
\bea
&&\la x_l(t) v_m(t) \ra = \int_{-\infty}^\infty d \Omega_1 \int_{-\infty}^\infty d
\Omega_2 ~(-i \Omega_2) ~e^{-i (\Omega_1+\Omega_2) t} \nn \\ 
&& \times \big[~ \cG_{l1}^+(\Omega_1)  \cG_{m1}^+(\Omega_2)~ \la
  \tilde{\eta}_{L}(\Omega_1)   \tilde{\eta}_{L}(\Omega_2) \ra \nn \\
& +& \cG_{lN}^+(\Omega_1) \cG_{mN}^+(\Omega_2)~ \la
  \tilde{\eta}_{R}(\Omega_1)   \tilde{\eta}_{R}(\Omega_2) \ra~ \big]
\eea
and this is readily evaluated using the noise properties in
Eq.~(\ref{noisew}).   After some simplifications we finally obtain:
\bea
G_l(\omega)&=&\big|~  \frac{1}{4 \pi} \int_{-\infty}^\infty d \Omega
~\Omega \nn \\
&\times& \big[ \gamma_L \big\{  \cG^+_{l,1}(\Omega -\om)-  \cG^+_{l+1,1}(\Omega
  -\om) \big \} \nn \\
&&~~~~~~\times  \big\{ \cG^+_{l,1}(-\Omega)+  \cG^+_{l+1,1}(-\Omega) \big\} \nn \\
&-& \gamma_R \big\{  \cG^+_{l,N}(\Omega -\om)-  \cG^+_{l+1,N}(\Omega -\om) \big\}
  \nn \\ 
&&~~~~~~\times \big\{ \cG^+_{l,N}(-\Omega)+  \cG^+_{l+1,N}(-\Omega) \big\} \big]
~\big|~. \label{exresp}
\eea
For nearest neighbor interactions, the force matrix $\Phi$ is a tri-diagonal
matrix. Using the properties of inverse of a tri-diagonal matrix we can 
explicitly evaluate the Green's function elements that are
required. For simplicity consider the case $k=1$ and
$\gamma_L=\gamma_R=\gamma$. Let us   define $\Delta_{l,m}$ as the
determinant of the sub-matrix of $[-M \Omega^2 +\Phi-\Sigma^+]$ that
starts from the $l^{\rm th}$ row and column and ends in the $m^{\rm
  th}$ row and column. We also define $D_{l,m}$ as the determinant of the
sub-matrix of $[-M \Omega^2 +\Phi]$ starting from the $l^{\rm th}$ row
and column and ending in the $m^{\rm   th}$ row and column. In terms
of these one has:
\bea
\cG^+_{l,1}(\Omega) &=& \frac{\Delta_{l+1,N}}{\Delta_{1,N}},~~\cG^+_{l,N}(\Omega)=
\frac{\Delta_{1,l-1}}{\Delta_{1,N}} \nn
\eea
with 
\bea
\Delta_{1,l-1} &=&D_{1,l-1}-i \Omega \gamma D_{2,l-1} \nn \\
\Delta_{l+1,N}&=&D_{l+1,N}-i \Omega \gamma D_{l+1,N-1} \nn \\
\Delta_{1,N}&=&D_{1,N}-i \Omega \gamma (D_{1,N-1}+D_{2,N})- \Omega^2 \gamma^2
D_{2,N-1}~.\nn 
\eea
For an ordered harmonic chain with all masses equal to $1$ it is easy to show
that $D_{l,m}=\sin(m-l+2)q/\sin q$ where $\Omega^2= 2(1-\cos q)$. Using this it
is easy to numerically evaluate the response function $G_l(\om)$ in
Eq.~(\ref{exresp}) for given
values of $l,N$. We show some numerical results in
Figs.~(\ref{harmcw_1},\ref{harmcw_x}) where we have also compared with results
from simulations for the linear response. As expected the exact response and
the linear response give almost identical results. However we have not been
able to analytically show the equivalence of the exact  response and the
linear response expressions.        
 
For large $\Omega$, we have $q \sim \pi+i \ln \Omega^2 $, hence 
\begin{equation}
\cG^+_{l,1}(\Omega) \sim 1/(-\Omega^2)^l. 
\label{cG_asymp}
\end{equation}
This can also be seen from the equations of motion: when $\Omega
>> 0,$ the dynamical equations become $-m_l \Omega^2 x_l = k x_{l-1}.$
The 
boundary condition is $-m_1 \Omega^2 x_1 = \eta_L(\Omega),$ in which
the right hand sign is effectively unity when calculating the Green's
function. Combining these equations, we obtain Eq.~(\ref{cG_asymp}).
But a $\sim 1/(-\Omega^2)^l$ dependence at large frequencies implies
that the $2l$'th derivative of $\cG^+_{l,1}(t)$ has a $\delta$-function at
the origin, i.e. $\cG^+_{l,1}(t)\sim t^{2l - 1}$ for $t\gtrsim 0$ (This can
be verified directly in the time domain: $x_l(t)\propto t^{2l - 1}$
satisfies the equations of motion for $t\gtrsim 0.$) But then in
the time domain, Eq.~(\ref{exresp}) is equivalent to $G_l(t) \propto
\cG^+_{l,1}(t) \partial_t \cG^+_{l,1}(t)$ for $t\gtrsim 0,$ where we
have assumed that $l$ is in the left half of the chain. Therefore
$G_l(t\gtrsim 0)\propto t^{4l - 3}.$ Since $G_l(t < 0) = 0,$ the
$4l - 2$'th derivative of $G_l(t)$ has a $\delta$-function at $t=0,$
so that 
\begin{equation}
G_l(\omega) \sim 1/\omega^{4l - 2}\qquad l < N/2
\end{equation}
for large $\omega.$

\section{Discussion}

In this paper, we have given an exact linear response formula for
the current in a wire in response to time-dependent temperatures
applied at the boundaries.  For a harmonic chain the full response function
has been analytically computed. We have presented numerical results for
the frequency dependence of the current response in oscillator
chains. For a diffusive system we find that the response differs
from what is expected from a solution of the Fourier's equation
with oscillating boundary temperatures.  It is straightforward to
generalize the derivation to fluid systems, various stochastic and
deterministic baths, and arbitrary system size $L$ and spatial
dimension $d.$ This is discussed in detail in Ref.~\cite{kundu09}
for the $\omega=0$ case.

As shown in Ref.~\cite{kundu09} the zero frequency response can be
expressed in terms of current auto-correlation functions, resulting
in an expression similar to the standard Green-Kubo formula but
without the thermodynamic limit being taken. If the integral of the
auto-correlation function remains finite in the thermodynamic limit,
the conductance is $\sim 1/N$ for large $N,$ and one can define an
$N$-independent conductivity in the same regime. The resultant
expression matches the standard Green-Kubo formula, but with the
thermodynamic limit taken {\it after\/} the range of the integral
is taken to infinity. While it is plausible to assume that the order
of limits commutes and 
\begin{equation}
\lim_{L\rightarrow \infty} \frac{1}{L}\lim_{t_0\rightarrow\infty} \int_0^{t_0}
C_{JJ}(t) dt =
\lim_{t_0\rightarrow \infty} \lim_{L\rightarrow\infty} \frac{1}{L}\int_0^{t_0}
C_{JJ}(t) dt,
\end{equation}
this is by no means trivial: if different boundary conditions had
been employed, with hard wall boundaries instead of heat baths, the
left hand side of this equation is zero but the right hand side is
not.  If the left hand side (with heat bath boundary conditions)
diverges in the thermodynamic limit, as for integrable systems or
low dimensional momentum conserving systems, the conductivity also
diverges, and one can only talk about the conductance or an
$L$-dependent conductivity.

At non-zero frequencies, the integral converges even when it does
not at $\omega = 0,$ and changing the order of limits is more benign.
Unfortunately, as we have seen in this paper, the expression obtained
for the finite frequency conductance involves the correlation
function $\langle j_{l+1,l}(\tau) J_{fp}(0)\rangle,$ which we are
unable to convert into an auto-correlation function when $\omega\neq
0.$ The connection to proposed expressions for the finite frequency
conductivity~\cite{bss06,volz01} is not clear.

\end{document}